# Arithmetic Deduction Model for High Performance Computing: A Comparative Exploration of Computational Models Paradigms

**Patrick Mukala**
Departement of Computer Science, University of Pisa

*Corresponding Author:*

Patrick Mukala,
Departement of Computer Science,
University of Pisa,
Largo Pontecorvo 5, 56127 Pisa PI
Email:patrick.mukala@gmail.com, mukala@di.unipi.it

## 1. INTRODUCTION

In [1] and [2], we succinctly introduced the Arithmetic Deduction Model (AriDeM) and present some results in light with a theoretical and empirical investigations conducted in comparison with the Von Neuman Model. The results indicated that with regard to number of instructions executed as well as execution times scaling through an exponential increase of in numbers of processors, AriDeM shows some indications of better performance compared to other paradigms. Admittedly with more experiments to be done, the initial results are indicative of its potential over existing paradigms. However, as we introduced the model, little has been said in the context of its genesis compared to most known models. Hence, in this paper we put all together and conduct an explorative study of these models in the context of requirements for high perfoamnce compting.

Considering the physical as well as logical configuration of computational machines, performance can be quantified and measured in terms of the speed and processing load of the processors. The architectural designs and processing mechanisms behind computational machines categorise the latter in two main processing modes: serial and distributed or parallel computation. In serial computation, data processing and instruction execution take place within a single Central Processing Unit (CPU) where a computational task goes through a number of processing steps. These steps involve the task being broken down first into isolated series of instructions, which are then executed one by one with only a single instruction executing at a particular point in time towards solving the computational task [40],[8]. This approach is based upon the famous von Neumann model, explained in detail later in this paper. On the other hand, the second approach entails the use of more than one processor to perform computation. This is parallelism and it is further elaborated throughout the rest of this literature survey. Many people see parallelism as an alternative to achieve high performance computing [12], [19], [61]. Silva and Buyya argue that parallel processing is the future direction in the quest for improved computing performance [56].

Behind the architectural designs in these processing modes, there is an appropriate model of computation. The von Neumann model is used in serial computing for architectural design and has been used successfully for a long time [56]. However, in parallel computing, this is still a major challenge. Unlike the von Neumann model that has been used in sequential computing for decades, there has not yet been such a model in parallel computing. Many parallel architectural models have been developed for this purpose, but none has proved to incorporate all the necessary simplistic and efficient properties required in a model of parallel computation.

For more than three decades now, computer scientists have tried to find an appropriate and more successful model of parallel computation in this regard, but with limited success. Some of the proposed models include the PRAM model (parallel version of the RAM model) [46],[5],[4], the Data Flow model, the BSP model, and the Network model [48],[24],[36]. While these models have played their part in the design of computational architectures, they have not all been entirely successful. This is due to a dynamic technological environment that continuously changes, as well as other factors. We shall explore some of these models of parallel computation. One of these models, the Parallel Random Access Model (PRAM), for example, has undergone a number of refinements, but these have not reached the ultimate performance for a successful model of parallel computation. This architectural model has thus proved to be unable to efficiently provide high performance results as expected [46],[64],[32]. As a result of this, it is critical to look for another alternative model of parallel computation, maybe even to consider another route other than instruction-based.

Therefore, the rest of the paper is structured as follows: in section 2 we give a short review of the use of parallel processing to achieve high performance; section 3 explores some models of parallel computation; in section 4 we highlight the need for an alternative model of parallel computation and conclude the paper in section 5.

## 2. PARALLEL PROCESSING FOR HIGH PERFORMANCE
### 2.1 State-of-The-Art Exigencies and Challenges

More advanced applications require more and more high performance computation. Such applications include engineering and scientific simulations, weather modelling, image and signal processing, as well as highly sensitive data retrieval. Some of these applications need very high performance reaching up to teraflops [46]. On a practical level, it is difficult to obtain such performance with serial computing. For such applications and any computational tasks that require very high performance, parallel processing has been considered an alternative in order to achieve the required performance [12],[19],[61], [56].

What is parallel processing? According to the Wisegeek Dictionary, parallel processing is the simultaneous processing of the same task on two or more microprocessors in order to obtain faster results. The computer resources can include a single computer with multiple processors, or a number of computers connected by a network, or a combination of both. Some supercomputer parallel processing systems have hundreds of thousands of microprocessors [71]. In addition to providing high performance computation, parallel processing can also be considered as an alternative to reduce cost. In [40], Kumm and Lea argue that the cost in terms of time and effort is higher when one attempts to perform complex computations using sequential computing than parallel processing.

Parallel processing offers quite a number of advantages that include: the reduction of execution time, the ability to solve bigger and more complex problems more effectively than using sequential computing and the ability to provide and support concurrency and more effective use of non-local resources.

The first advantage is execution time reduction. As more and more advanced applications are developed, the focus is also on obtaining results faster and in reduced time [73], [17]. Because of the constant need to reduce the amount of time taken to solve a computational task, parallel processing is deemed appropriate to accomplish such results. Coffey and colleagues acknowledge that computations previously taking many days with traditional mainframes are now being performed in only a few hours, thanks to parallel processing [22]. It is important to point out at this stage that the amount of time that an executing instruction takes is an important factor that determines the performance of a particular architecture [46]. The ultimate architecture is required to produce results in reduced time while minimising communication costs [46].

Moreover, the time requirement for parallel and sequential computing evaluation works hand in hand with the speed [40]. This is important in evaluating the performance of both the AriDeM and von Neumann models. How efficient is the execution of the same computation on both approaches? Which one is more efficient and cost-effective? This constitutes another important issue that needs to be answered in this research for both the von Neumann and AriDeM model. The speed used by each of the two computational architectures to perform the same task is important in determining which architecture is likely to perform better than the other [40].

Another advantage of parallel processing is its ability to allocate multiple resources to handle complex and bigger computational problems. Sunderam [59] supports that very complex problems requiring high performance computation such as image processing, protein folding, genomics and Core-Collapse Super-Nova [34] require parallel processing. Another problem that can be solved effectively using parallel processing is the problem of computational protein engineering in genetics, in which an algorithm is sought that will search among the vast number of possible amino acid sequences for a protein with specified properties. Yet another example is searching for a set of rules or equations that will predict the ups and downs of a financial market, such as that for a foreign currency [49]. Such problems can be solved efficiently by making use of parallel processing, because many different possibilities can be explored simultaneously to get the results. For example, in searching for proteins with specified properties, rather than evaluating one amino acid sequence at a time, it would be much faster to evaluate many simultaneously. Other problems include groundwater modelling [62], digital signal processing and systems for health care as well as economic change [55]. More complex areas such as stornadic storms in order to explore in detail and examine the intricate interplay of temperature, moisture, turbulence, air pressure and wind in the genesis and development over time of a full-scale twister could only be done using parallel processing [55].

Although parallel processing gives a wide range of benefits, there are some challenges that are still to be overcome. The ultimate challenge that is the subject of our research is an appropriate model. Other challenges exist as well. Implementing parallelism in computing is not an easy task. One challenge is that the complexity of parallel, networked platforms and highly parallel and distributed systems is rising dramatically [55]. Today's 1,000-processor parallel computing systems will rapidly evolve into the 100,000-processor systems of tomorrow [55]. However, the existing architectures can only provide limited results in this regard [12]. Hence, another challenge in parallel computing today is software that is scalable at all hardware levels (processor, node, and system) [55], [47]. Achieving scalability is a major challenge in parallel computation and comes with a cost [58]. Do existing models of parallel computation balance all these factors (scalability, communication cost) to achieve high performance? Current results suggest that this has been a major challenge for some of these models.

Another challenge in parallel architectures today is the inability to exploit multiple hardware levels successfully [55]. Although parallelism in computation is of the utmost importance, computational science also requires scalability in other system resources. For example, to exploit parallelism in memory architectures, software must configure communication paths to avoid bottlenecks. Similarly, parallelism in the I/O structure allows the system to hide the long latency of disk reads and writes and increase effective bandwidth, but only if the software can appropriately batch requests. Existing models have not been able to accomplish this more successfully [8].

**2.2 Examples of Parallel Computation Applications**

In the United States of America, as well as in other areas of Western Europe, the use of parallel computing has made headlines in both the private and public sectors. A great deal of high performance systems in many areas that are currently in use have been designed using parallelism. Let us look at some areas where parallel processing has made an impact.

Given the performance resulting from distributed computing, new methods for collecting and analysing data have enabled the social and behavioural sciences to record more and more information about human social interactions, individual psychology and human biology. In the US, rich data sources that can be managed through parallel computation include national censuses, map-making, psycho-physical comparison, survey research, field archaeology, national income accounts, audio and video recording, functional magnetic resonance imaging etc [10],[40].

High-profile problems such as the US DARPA grand challenge and many more projects in engineering and science carried out to develop applications with high-performance computing resources have succeeded because of parallel computing [58]. Some of these projects include tackling problems in computational fluid dynamics to simulate the design of more efficient and robust aircraft and extremely silent submarines to enforce military power [58].

Another accomplishment of parallel processing is faster Internet applications. With the fast and ever-growing size of the Internet, there are millions and even billions of database transactions that need to be efficiently processed [42]. In the same environment, there are other additional constraints that are to be taken into consideration, depending on the nature of the executing applications. Some of these constraints might involve large amounts of real time data that must be processed quickly, while others are just very complex

[58]. Processing a huge number of transactions and delivering the results in due time, as well as overcoming all the transaction-related constraints is the realm of parallel processing [58].

Another tangible example of parallel computing application is ACCESS GRID. This is a global network, which provides virtual collaboration around the globe. It groups a considerable number of people to work on similar tasks at the same time [21]. The SETI@home project [10] is another parallel application. This application enables an effective utilisation of remote resources. Not only does it allow for communication between these resources, but it also enables their appropriate coordination to allow a more efficient and concurrent task execution to provide network effectiveness [10].

In SETI@home, resources from millions of computers around the world are used in search of extra-terrestrial intelligence, thus performing the largest computation ever [10]. During this process, radio signals from space are analysed on millions of computers used either at work or at home around the world and this effort may provide insight into what is needed for extra-terrestrial intelligence predictions [10]. While the process seems like a very complicated approach, and it is indeed a very complex endeavour, the technological impact is made, thanks to the power of parallel computing.

These selected examples exemplify the use of parallel processing in this research. In the next section, we do a review of some models of parallel computation

## 3. MODELS OF PARALLEL COMPUTATION

Models of parallel computation have been the realm of many research studies for years [46], [47], [57]. With numerous refinements and improvements, some of these models have been fine-tuned in the quest for a more appropriate model of computation of parallel computation. Effective computational models are computer architectural models that incorporate properties that support scalability, portability and performance, while reducing cost [12], [59]. These are important factors that contribute towards developing efficient architectures. A successful model lays the foundation and provides a unifying platform for both algorithm developers and hardware designers to develop parallel applications [59]. The von Neumann model is regarded as successful also, because it provides such a unifying platform.

One of the challenges of parallel computing today is the existence of a simple model of computation. Some of the existing models have been of limited success because of their complexities. Lee argues that an effective model of computation should be simple. On such, modelling all complexities with regard to distribution and coordination of instructions across multiple processors should be simplified [42].

Why the need of a model and what is the whole point of modelling? Before discussing the different proposed models of computation, we believe it is important to describe briefly modelling and its significance. Researchers in various areas of technology and science have constructed and made use of models for many years to help understand and test theories [50]. According to Moor, modelling is an important and major activity in computing. Basically, models in computer science are developed to simulate computer hardware or software designs and they give an indication of the entire process - everything from the contraction of the heart muscle to the growth of rice crops [50]. In computing, a model is much more than just a computer program; it is a representation of the significance of a real event that is a real product or action. Additionally, architectural as well as computational models are core factors in both machine and software designs. The successful design of computing resources and architectures critically depends on the relevant model [47].

There exist numerous models of parallel computation. However, in this section we look only at some of these existing models, which include the PRAM model, the DataFlow model and the BSP model, as well as the Multicore approach.

### 3.1. Parallel Acces Machine (PRAM) Model

In 1945, John von Neumann authored the general requirements for the design of a computer. He came up with an architectural model known as the von Neumann Architecture. This architecture comprises four main components: Memory, Control unit, Arithmetic Logic and Input/output. The following paragraphs give a brief description of each of these components.

In this architecture, random access memory is used as part of the memory component to store both instructions and data. These instructions depict coded data on how the computer operates; they communicate to the computer what to do. The data in memory represent information that programs make use of to execute instructions [24], [41]. The control unit performs the coordination tasks by fetching information from the memory (instructions and data), decoding these instructions and then coordinates in sequence the actual operations to accomplish a programmed task. The Arithmetic Logic unit, on the other hand, performs any operations in relation to mathematical or arithmetical requirements during the undergoing sequential instructions execution, while the Input/output is the component that constitutes the interface with the computer user [41]. Warren states that in this architecture, the core functionality is with the processing unit associated with a stored program, as well as repeated cycles of fetching and executing instructions [69].

It is worth noting that, despite the successful dominance and impact that the von Neumann computation model has had over the years in serial computing, increased performance has always been a concern for computer scientists. Improving the processing time and speed of instructions execution, as well as producing more powerful machines constituted, among others, critical milestones that researchers were concerned about. Due to this success, the PRAM model was developed for parallel computing, in the hope of obtaining the same results. Thus, to improve the performance of von Neumann-based computers, some innovative improvements were made, as discussed below.

The first improvement that is worth mentioning here came in 1949 from a group of researchers at Manchester University. This is the Indexed modification of addresses and the memory hierarchy. This architectural modification ensures that the index registers would permit the execution of loops with no modification effect on the instruction addresses, and programs would be automatically allocated in memory. This concept of memory hierarchy was the basis for the consideration of caches and virtual machines in sequential computing [24]. Another breakthrough came two years later around 1951 with the introduction of multi-programmed control by Wilkes [36],[41], [16]. This control mechanism was intended to control the operation of computers in a new and more systematic way. The effects of this research on the subsequent machines have been tremendous [16].

Next, the stack architecture was introduced as a result of emerging innovations at the heart of the von Neumann-based architecture. Developed in 1958 by Barton, it was considered as a tool that would be used in order to improve the compilation and execution of expressions [24]. This configuration at the machine level carries significant advantages, including the necessity of reflecting a programming language organisational, the efficient management of subroutine invocations for operating systems, as well as general program context.

Another significant improvement introduced was the use of the multiprocessor concept. First introduced in the fifties, this brought about changes with the concept of separate I/O processors, as well as the arithmetic processors to improve computer performance [24]. Another innovation came along with the introduction of the pipeline architecture that was also used in the vector arithmetic processors to improve machine operations on data structures [24]. The pipeline architecture helped get improved results through decomposing operations in steps that cascaded subunits will execute.

A last innovation to be mentioned in this research stemmed from the pipelined architecture. A different architecture developed on the basis of the pipeline architecture is called the systolic arrays architecture [36], [41]. This architecture was characterised by the presence of processing elements such as programmed processors that are identical and connected in a linear or multi-dimensional array [24]. This configuration is set up in such a way that each processing element is connected to its adjacent elements only. These architectures have been made use of in dedicated VLSI components and other high-performance computers used for special purposes such as images and signals processing. Other research studies name the increase in the length of a single word in a processor as another improvement made to von Neumann model-based architectures [69].

However, in the middle of all these changes and especially with the introduction of microprocessors and the improvement of component density, studies demonstrate that this has resulted in the speed of involved microprocessors being relatively slow because of the complexity linked to instructions set [24]. Furthermore, the use of these complex instruction sets was rarely effective in programs and, because of these results, there had been an increasing need to consider going back to designing architectures that would be based on a relatively smaller number of instructions to implement with simpler circuitry and therefore would work at a considerably higher instruction rate [24].

Based on the performance improvements discussed above, one will note that the need for better computer performance has been the realm of research since the early days of computing. These innovations of computer architectures, along with innovations in other computing areas such as hardware components as well as software and application technologies, have contributed significantly to major refinements and improvements in computing over the years [12], [48], and [46].

The quest to implement parallelism in existing computer architectures could be seen in two different tendencies [24]. The first direction was motivated by the need to create new architectures as a means to pave the way for the expected increased computing power. This tendency would start its experimentation by exploring new models of computation that would be drastically different from the classical serial model [24].The second direction would be tackled using a dual approach. This means that it would consider the extension of the von Neumann architecture to include other features such as parallelism that would be implemented on sequential machines operating in parallel on cooperating processes [24]. Thus, new architectures were created and would be referred to as "parallel von Neumann architectures" and these architectures were based on the PRAM model.

Some machines, including the very first ones that were built based on these architectures, implemented a configuration with one master unit connected to several other units called slaves and the latter operating under the total control of the master unit [24]. This architecture at some extent made it relatively easy to solve the problem of synchronisation. However, another even more difficult problem in this regard was related to distributed control, each machine interacting with the other, all at the same level [24]. The solution to this problem was the design of specially conceived circuits capable of handling interrupts as well as refinements such as semaphores brought to programming languages to handle synchronisation.

The parallel random access machine (PRAM) model was developed in parallel computing in order to emulate the success of the RAM model in serial computing [44],[4]. The motivation was to create a corresponding parallel version of the von Neumann model due to its success in maintaining consistency and coordination among software developers, algorithms designers as well as architects in serial computing [46], [48]. This model consists of a set or grouping of processors that execute the same program using a shared memory in a lockstep fashion. Each processor is able to run or execute its allocated instructions and access a memory location in one time step, independent of the memory location.

According to McColl in his research on parallel computing, a parallel random access machine is one that consists of a collection of processors configured in such a way that computation occurs synchronously in parallel and these processors communicate via a common global random access memory [48]. In addition, in some PRAM variants, as explained later in this study, processors can only communicate by writing to, and reading from, the common memory. These processors have indeed no local memory other than a small fixed number of registers which they make use of to store values temporarily [48].

In some tasks, the PRAM architectural model has been able to allow the designer to exploit the possible computational parallelism to the fullest, despite associated costs [46]. However, in order to provide high performance while reducing cost, the PRAM model needed refinements based on a number of factors. These factors include memory access, synchronisation, the processing speed or latency, as well as the data transfer rate associated with every transaction or bandwidth [48], [46]. The need to improve these characteristics in the PRAM model led to the development of its many variants, as explained below.

Memory access in the PRAM models varied on specifications, depending on the implementation of its variants. Some variants such as the CRCW (concurrent reading or concurrent writing) PRAM embeds flexibility with regard to concurrent writing or concurrent reading at any memory location, through control properties such as prioritising as well as randomisation or a combination of the two to ensure an effective concurrent writing [46]. Providing and controlling concurrent memory access at any memory location within this variant of the PRAM model comes at a considerable and unavoidable cost. But this needs to be alleviated, as the ultimate outcome of parallel computing is to ensure improved performance within a very cost-effective context ([44]; [4]; [48]). This implies reaching high performance processing at reduced cost throughout the entire instructions execution process. For this reason, the PRAM model was once again refined to develop more variants to try and alleviate the issue of cost linked to the concurrent memory access property of the CRCW PRAM. One of these variants is the CREW or EREW PRAM which puts restrictions concerning memory access. These restrictions authorise only single memory access of processors one at a time for either reading or writing or even both [46]. Hence, the PRAM model brings some change for an effective use of concurrent processing and memory access, but fails to enable a strict and efficient parallel computation approach, as it enforces serial processing [46].

Another PRAM model variant called the Module Parallel Computer was subsequently developed with the aim of improving the previous models and hopefully producing a more efficient architectural model for the purpose of conducting the design of parallel products [46],[44].This model uses a shared global memory structure organised in such a way that the global memory is broken down into a number of modules, with each memory module allowing a single memory access based on a specific time frame. However, this PRAM model variant still does not solve the entire problem as such, by providing a significant concurrent memory access and processing approach, but it alleviates the serial processing and access at the global memory encountered in the previous variants.

Also in the same line of thought concerning memory access in PRAM models, one last variant of the parallel random access machine that was developed once again in order to improve the previous variants and, more importantly, to resolve the mystery behind an effective memory access for capitalised parallel computing performance is the Queue-Read, Queue-Write PRAM (QRQW) parallel random access machine. In this PRAM variant, another means of controlling concurrent memory access is inserted differently from the previous variants. The model introduces a queue as an intermediary means of coordinating access to memory location throughout the entire parallel computing instructions execution [46]. But the use of queue to ensure a proper memory access is not an optimal solution, as it is costly to the performance of the entire architecture in terms of performance delay due to the relative length of the queue.

One can notice from the development of these PRAM variants that memory access has been a very critical issue to deal with in these models. This is also a major challenge that the new required computational model is expected to solve. Implementing successfully an appropriate memory access mechanism in the PRAM models always came with a cost on performance, as other functionalities were inextricably affected in these models.

Furthermore, the design of parallel random access machine models also encountered a problem with regard to the synchronisation of concurrent instructions execution. In the above-discussed PRAM variants, the synchronisation of all involved processors throughout the entire instructions execution occurs through a global clock. However, some PRAM variants such as the XPRAM employ a different approach to ensure a prominent synchronisation mechanism during the instructions execution [46][12]. While bypassing this restriction, these variants use the asynchronous execution mechanism, hence forming a class of APRAM models or Asynchronous Parallel Random Access Machine. In these models, however, such as the XPRAM as indicated above, there is still synchronisation incorporated and occurring on a periodic basis between specific intervals throughout the entire asynchronous execution process. While these models incorporate synchronisation, they do not charge an explicit cost [44]. In this group of models, the impact on the expected ultimate performance is not as significant as it is in the PRAM models [44]. This impact relates, for example, to the processing delay due to the loss in the use of specific processor while waiting for the response from the other processors still executing instructions. Although the only cost is implicit with these models (the loss of processor utilisation while waiting for other processors to complete), these models still provide an incentive to synchronise only when necessary.

Synchronisation constitutes a very important and critical factor that requires proper implementation and execution between the different processors performing the concurrent execution of instructions in parallel computing. It is vital that instructions be successfully coordinated and synchronised during data processing, in order to ensure efficient outcome within an ultimate and reduced period and, very importantly. in a cost-effective fashion. However, many PRAM variants have not been able to tackle this design threshold without affecting the overall architectural performance.

In addition to the parallel computation characteristics discussed above, there is another non-negligible factor that requires special attention in the development of any given parallel and distributed computing architectural models: latency [46], [44], [4]. A latent tasks execution is very critical, as it affects not only the overall design performance but also subsequently has negative effects on the productivity of the users. Perceived as a symptom of low performance in serial computing, latency makes integral part of the motivation behind the quest for high performance computing. Built upon fewer resources for distributed and parallel processing, serial computing could not handle large chunks of data or instructions to be processed in a cost-effective and timely fashion; hence, this would result in either the processing applications being extremely slow or other performance abnormalities [46].

In order to avoid latency in instructions execution, other variants of the PRAM model were developed to reduce the cost involved from the use of non-local memory access. The first variant is the LPRAM model, where compromises were done at the local memory access in order to access the global memory [4]. Another PRAM variant that came into light concerning reducing the problem of latency in parallel computing was the BPRAM model [46]. In this architectural model, there is an emphasis on data parallelism, whereby the communication costs related to memory access occur first at the local memory level and then from the global memory for each message or block transfer that takes place [5].

In order to alleviate the latent communication issue in parallel computing using the BPRAM model, Aggarwal and his colleagues created this PRAM variant by suggesting a different approach. In this approach, global memory is not used as in previous designs, but the global memory is segmented into small locations that can be accessed at specific times and, in this approach, processors establish a communication link by exchanging messages. In addition, a processor can only receive or send one message at a time [5]. Other communication parameters such as bandwidth, synchronisation requirements, as well as the problem of congestion in the communication network, are part of simplifications that come with the BPRAM models [5].

Furthermore, the BPRAM model was one of the most utilised models developed towards the end of the 1980s, right before the existence of the current massive architectural models. In the current existing parallel computational models, the communication costs between processor clusters as part of parallel architecture are very high and are relatively linked to the bandwidth [46]. However, the BPRAM model has very significant properties, allowing the design of efficient parallel algorithms as well as other computational designs requiring parallel computing.

Another factor that drove the improvement of the PRAM model is bandwidth. In the parallel random-access machine, communication takes place on an established network where information is conveyed between processors and the available memory [44]. The assumption concerning concurrent processing or parallel communication revolves around the fashion in which each of the clustered processors accesses the memory location. It is assumed that in the different PRAM model variants, each processor is able to access a single memory location simultaneously based on a specific time scale. However, in very large parallel computers, this assumption would require implementation of networks where communication bandwidth is a very demanding and key factor. Nevertheless, the implementation of such networks based on this assumption is difficult [44].

Hence, another variant of the PRAM called distributed random-access machine or DRAM was developed. This model is based on a more restricted approach in terms of the above assumption, compared to other previous PRAM model variants. The distributed random-access memory model (DRAM) was built around the approach where, in a parallel computer, memory accesses are implemented in such a way that messages are routed through a communication network. This model design enables the modelling of problems related to messages congestion across the underlying network segments [44]. The DRAM model is invaluable in this regard in the sense that its configuration design is based on a very particular approach that eliminates the pattern of global shared memory encountered in the majority of the other PRAM variants discussed above [46]. It caters for the global memory by replacing this configuration with a distributed memory approach. The communication network topology does not have significant impact on the entire communication process and the DRAM incorporates an approach that only requires limited communication bandwidth [46].

In the DRAM model, all memory is local to the set of clustered processors and each processor holds a small number of registers of a bit magnitude [44]. In this model, a processor uses values stored in its local memory in order to perform operations such as read and write, as well as logical and arithmetic functions [44]. Furthermore, the DRAM model allows each clustered processor to read and write memory in other clustered processors. It can additionally use the local temporaries to send data between remote memory locations [44].

These few paragraphs have attempted to review the different alterations and refinements that were brought to the different variants of the PRAM model in the quest for a better alternative to parallel computational and architectural models. Although the ultimate parallel architectural model was not built throughout this model, the current generation of parallel architectures does in fact reflect these characteristics expressed, because of modifications brought about to the PRAM model.

Lastly, in his studies on parallel computing, McColl argues that one of the primary objectives of the PRAM model since its development was to facilitate the evaluation and study of parallel algorithms as well as associated complexities [48]. However, as technology evolves and triggers the need for parallel architectures implementation, studies have shown that there is no framework for the design and programming of such architectures. Thus, there is an imminent concern to work on a new model that will cater for all these imbalances.

### 3.2. DataFlow Model

The Data Flow model is one of the paradigms that were introduced as an alternative to the parallel von Neumann computation model [48] and [70]. This model was also designed to effectively express parallelism. As the name suggests, the Data Flow model is a parallel computational model that is data-driven. This means that the sequence in which the execution of instructions occurs totally depends on the availability of operands that are needed by the corresponding instruction, while unrelated instructions are executed concurrently without interference [70]. According to Arthur Veen, data-driven also means that in order for a process to be activated, its input data needs to be available, because this activation is solely determined by the availability of the former [67].

The data flow architectural model emerged in the 1970s. It is based on the concept of dataflow program graphs to explore parallelism in programs [11], [25]. In the model, operations are determined by actors that are activated or empowered when all relevant actors producing data needed, have completed their execution [25]. Furthermore, graphs are used in this model to represent the dependence between pairs of actors, where a message is conveyed from an actor to another called or referred to as a successor actor. These graphs also show the different firing rules that represent and specify the type and nature of data needed as a requirement for an actor to fire or start an execution [67]. Additional actors can also be added to represent iterations as well as conditional expressions.

A data flow graph represents the essence of the incurred computation. The graphs used in the data flow model for representation purposes have two main characteristics: the involved data is represented by arcs or values and not by memory location; occurring operations are functional, meaning that the produced output totally depends on the input [52].

Mainly four advantages can be observed with using the data flow model. The order in which operations occur is determined by true data dependencies and there are no constraints imposed by artificial dependencies that could be introduced by stored program model including memory alias etc [52]. The model represents parallelism at all levels during computation. This implies that at the operation level, there is fine-grained parallelism displayed and this can be exploited within a processor, function or loop level. Parallelism can also be exploited across processors while systolic arrays exploit only fine grain parallelism [52]. At the semantic level, combinational logic circuits are naturally functional, meaning that the output depends on the inputs and as a result these circuits are easily synthesisable from graphs in the data flow diagram. Finally, an additional advantage is the lack of implicit state in dataflow graphs, and therefore they allow easy and correct synchronisation of sequential circuits where the storage elements are determined by the synchronisation tool rather than by a compiler [52].

Furthermore, one of the other advantages that this model has over the von Neumann architectural model includes, for example, the ability to significantly reduce the complexity of connection between processors and memory encountered in von Neumann models [70]. This complexity encountered in the von Neumann architectures can be noticed with the use of semaphores. In order to process recurring signals in these machines, a complex system of semaphores is required as a means to solve a computational problem. However, since the purpose of signal processing is to attain increased throughput and that the amount of data to be dealt with is relatively small some studies have demonstrated that the Data Flow computational model would be perfectly fit to handle such processing because it embodies an asynchronous execution of instructions [70].

Moreover, it is important to mention that as a parallel computation model, the dataflow approach was considered by many as the answer to overcome parallel processing problems that the pre-existing architectures could not solve. This model has had a remarkable influence on many areas of information technology and computer science, as well as research in engineering and science. Major influences of this paradigm can also be noted in products such as programming languages, processor design, multithreaded architectures, parallel compilation, high level logic design, signal processing, distributed computing, and programming of systolic and reconfigurable processors [52].

However, despite these results, research has proved that the impact of the dataflow architectural model does have some limitations. According to Professor Dennis from MIT, there were still problems to be addressed with the dataflow architectural model [27]. One of the limitations is related to the type of programs data flow computers can support. His research argues that it is difficult to see how data flow computers could support programs written in FORTRAN without restricting and carefully tailoring the used code [27]. Additionally, much research needs to be conducted in order to improve the problem of mapping high-level programs into machine-level programs that effectively make use of machine resources [27].

Iannucci argues that architectures based on the data flow parallel computational model do need improvement in terms of efficient toleration of latency and synchronisation costs [37]. Different optimisation and synchronisation techniques are implemented on diverse variants of the model. Such variants include, for example, the MIT Tagged Dataflow Architecture where synchronisation is achieved through a complex mechanism to synchronise the execution of instructions in proportion with time. Finally, another important constraint encountered in the data flow diagram is the use of a finite cyclic graph to represent a program. Due to evolving technological innovation and an emerging need for high performance computing, this will yield unlimited success.

### 3.3. Bulk-Synchronuous Parallel (BSP) Model

The Bulk-Synchronous Parallel (BSP) model is a model of parallel computation developed and proposed by Leslie Valiant as an interface or bridging model between parallel software and hardware [32]. As described by the author himself, the BSP model has been designed and aimed to allow the optimal execution of architecture-independent software on a diversified number of architectures [64],[32].In the BSP model, the configuration of a parallel computer consists of a set of processors associated with their own respective local memories, and a communication network that performs interconnection messages passing by routing packets in a point-to-point fashion between processors [48] and [32]. The operation or functioning of the model occurs as follows:

A computation in a BSP Model is divided into sequential supersteps. In this sequence of supersteps, each of these supersteps is also a sequence of steps followed by a barrier of synchronisation where all memory accesses occur [64]. During each superstep, a processor carries out a number of activities based on its data and programs. These activities include carrying out computational steps from its threads on values received at the beginning of the superstep. Each processor can also perform additional operations on its local data, can send and receive a number of messages and packets [64]. As further elucidated by [32], a packet sent by a processor during a superstep is delivered to the destination processor at the beginning of the next superstep [32]. The synchronisation barrier is used to separate successive supersteps for all processors (Ibid, 1999). Three fundamental parameters are used to model a parallel computer based on the BSP model. These parameters include the number of processors $p$, the interval or gap $g$ and latency $L$. The interval $g$ reflects network bandwidth on a per-processor basis, the latency $L$ depicts the minimum duration of a superstep and reflects the speed with which a packet is sent through the network, as well as the time taken for synchronisation [32]. This BSP configuration offers a couple of advantages, as argued by the author.

One of these advantages is related to the level of abstraction that the model offers. It offers both a powerful abstraction for computer architects and compiler writers, and a brief model of parallel program execution. This will enable a performance prediction that is accurate for proactive software design [64]. A simulated BSP abstract computer would consist of a set of virtually grouped processors associated with local memories on a communication network whose monitored properties include the speed or time for synchronisation and the rate at which continuous randomly addressed data can be delivered [32].

Furthermore, McColl argues that the BSP model helps lift the considerations of network from the local level to the global level. This describes the ability to support non-local memory accesses in a much more uniform and efficient fashion [48]. A further advantage of this model is its potential for automating memory management via hashing [48]. It also provides a framework that is independent of the architecture, which allows the designers, including computer architects, to take full advantage of whichever values of $L$ and $g$ are the most cost-effective at a given point in time.Computation in a BSP model consists of a sequence of parallel supersteps, during which computation and communication occur, and consecutive supersteps are separated by what is called a synchronisation barrier [32]. This configuration and functioning of the BSP model make it compatible to some extent with the conventional SPMD (single program, multiple data) model as well as the MPMD (multiple program, multiple data) model. The BSP model also incorporates both remote memory and message-passing capabilities (BSMP). However, the difference is clear in terms of the timing of communication operations; this is different since the effects of BSP communication operations do not become effective until the next superstep [32].

Other important features that are worth noting concerning the BSP model pertain to its capabilities in terms of portability and efficiency. As demonstrated throughout this survey, a very important objective or significant goal of parallel computation is developing parallel applications or programs that are portable and, most importantly, efficient. Valiant has argued that his Bulk-Synchronous Parallel (BSP) model achieves both portability and efficiency for a large class of problems. The cost of portability is that BSP code may require a larger input size than machine-specific code in order to achieve the desired level of efficiency [65].

Research on parallel architectures has demonstrated that scalability in parallel performance can be achieved for a given number of computational problems using software that is independent of the machine performing the execution [48], [64], [65], [32]. However, this has been a great challenge to achieve during the quest for a more robust architectural model of parallel computation. This is due to some extent to the fact that portability can be achieved at the expense of achieving scalable performance. On this note, Valiant supports that an effective way of handling this is by proposing and developing a unifying or bridging parallel architectural model that will serve as an interface between the software and the hardware; and as elaborated, he proposed this model, the Bulk-Synchronous Parallel (BSP) model, as a prospective candidate fulfilling these requirements [64].

Moreover, in supporting the effectiveness of the BSP model, Goudreau et al. argue that this model is so dynamic that many other programming styles can be transformed into a BSP style automatically and efficiently [32]. To support this, it is worth mentioning the different simulated refinements that were brought about to the model and that proved that the BSP model could efficiently simulate the EREW PRAM [32]. The result of this simulation was subsequently extended to the more powerful QRQW PRAM model by another group of researchers [30]. In addition to these experiments, another simulation of asynchronous message-passing programs in BSP was given by a different group of researchers in 1996 [18]. Another simulation was performed by Gibbons et al. in demonstrating the emulation of another different proposed model called the QSM (Queued Shared Memory) model [29].

Although the BSP model has shown some indications of supporting diverse platforms and programming styles by emulating different types of models, the result has yet to be the expected optimal model. There are still a lot of challenges to conquer within the BSP model itself, such as developing an appropriate programming framework for the model [48]. In this regard, few refinements and extensions of the BSP model developed with the purpose of attaining better performance can be noted. The resulting variants are the parallel-prefix computation (PPF-BSP), broadcasting (b-BSP), and concurrent reads and writes (c-BSP) [32].

### 3.4. Current Attempt: Multicore

In 2007, Chai et al. conducted a research on understanding this architectural platform in the light of the exponential factor elucidated by Moore's law. In that research study, they proved that it would be difficult to increase the speed of processors nowadays by just increasing frequency, as per Moore's law [20]. This is due in part to the problem of power consumption as well as the overheat barrier. Due to these factors affecting the performance of multiple processors increasing exponentially by increasing the processor clock rate, the architecture would become less effective in terms of cost concerning provided performance [20]. Therefore, the multicore paradigm was considered by computer architects in order to try to alleviate this problem to some extent.

The multicore processor (also referred to as Chip Multiprocessor or CMP) architecture means that two or more processing or computational cores are placed on the same chip. These processors increase the speed of applications performance by dividing the inherent workload to different cores [20]. As mentioned previously, one of the fundamental reasons motivating this approach is the fact that scaling up processor speed results in a dramatic rise in power consumption and heat generation, and thus this needed to be dealt with by incorporating multiple cores on a single chip to speed up the processing performance.

Moreover, Parkhurst and colleagues in explaining the migration from single core to multicore believe that parallelism is one the best ways to address the issue of power in single core, while maintaining performance where higher data throughput may be achieved with lower voltage and frequency [53]. The result is, however, a larger transistor count, but overall lower power dissipation and power density.

Furthermore, it is important to note that nowadays, it has become quite a difficult issue to try to increase processor speed because of associated costs. Taking all these factors into account and, most importantly, considering the need for better and faster architectural models, multicore processors have been proposed to speed up application performance by dividing the workload among multiple processing cores instead of using one super fast single processor.

Other advantages of this approach include inherent redundancy, which lends itself to resilient architectures. Instead of binning based upon speed, one could bin products based upon the number of working cores or overall data throughput. In addition, the integration of multiple cores on a chip allows for lower interconnect latency and therefore higher bandwidth between cores than their discrete counterparts do [53].

This paradigm has gained quite a considerable market in the computing industry and mostly for computer and hardware manufacturing giants. Some of these manufacturing companies' multicore products include Intel Quad- and Dual-Core Xeon, AMD Quad- and Dual- Core Opteron, Sun Microsystems UltraSPARC T1 with 8cores, IBM Cell, etc [53]. Multicore processors have been deployed and used in many computing areas, including cluster computing. In a multicore-based cluster, for example, Chai and colleagues [20] argue that there are three levels of communication that multicore makes possible in the cluster. The first level depicts the communication between two distinct processors placed on the same chip; the second level depicts the interconnection across different chips and the communication at the last level depicts the communication across chips but within a node [20].

Despite all these advantages, the paradigm still has some challenges. Multicore clusters do impose considerable challenges pertaining to software design, both at the middleware level and application level and another issue not to be overlooked relates to cache and memory contention. This is critical because in multicore clusters, it can be a potential bottleneck in multicore clusters, and in order to alleviate this problem, the middleware as well as involved applications should be programmed respecting these specifications [20].

In addition to these previous challenges, Parkhurst and colleagues argue that new challenges arise with respect to the non-core or glue logic of the design. They support that, with respect to caching in multicore, whether we assume that there is a dedicated cache per core or cache-sharing, the need to maintain cache coherency will create added complexity [53]. Furthermore, the authors' research sustains that in the near term, there is a possibility that there will not be enough transistors available to allow the implementation of distributed cache scenario, thus forcing cache sharing. In this case, complexity will increase as designs try to manage access as well as cache hits and misses.

Software is another very important aspect to be considered in this regard because of the prominent role it plays during this multicore era. As numbers of cores per chip will be increasing exponentially, there will be diminishing returns in this paradigm if software applications do not fully utilise the processing power at their disposal. Software's ability to efficiently partition and distribute workload among the cores will be the key enabler [53]. Moreover, Valiant also believes that in terms of software, the designer of parallel algorithms for multicore machines has to face so many challenges that the road to their efficient exploitation is not clearly signposted. Among these challenges, we consider the following: firstly, the underlying computational substrate is much more intricate than it is for conventional sequential computing and hence the design effort is much more onerous. Secondly, the resulting algorithms have to compete with and outperform existing sequential algorithms that are often much better understood and highly optimised. Third, the ultimate reward of all this effort is limited, at best a speedup of a constant factor, essentially the number of processors. Fourthly, machines differ, and therefore any speedups obtained for one machine may not translate to speedups on others, so that all the design effort may be substantially wasted. For all these reasons, Valiant

[66] argues that it is problematic how or whether efficient multi-core algorithms will be created and exploited in the near future.

Another challenge concerns the effectiveness of accommodating latency linked with communication between chips, both in protocol and simple interconnects. Chip architects generally decide very early in a design project's schedule exactly how a design is going to behave. They make these decisions based upon their past design experiences, new research discoveries, and feedback on implementation feasibility from silicon circuit designers.

To add on to the challenges associated with multicore, let us consider the Berkeley group's view about this matter. In their study on landscape of parallel computing research, [12] established that multicore would help multiprogrammed workloads, which contain a mix of independent sequential tasks [12]. However, the problem is to make individual tasks faster. They believe that migrating from sequential to parallel computing will consequently make programming much more difficult without rewarding this greater effort with a dramatic improvement in power-performance. Moreover, because of this analysis, they believe that multicore is unlikely to be the ideal answer. Hence, they would consider manycore rather than multicore. They substantiate this view by arguing that it is unwise to presume that multicore architectures and programming models suitable for 2 to 32 processors can incrementally evolve to serve manycore systems of 1000s of processors. Their research further elaborates that the aim is to realise thousands of processors on a chip for new applications, and this platform can be improved over time by new programming models and new architectures if they simplify the efficient programming of such highly parallel systems. Successful manycore architectures and supporting software technologies could reset microprocessor hardware and software roadmaps for the next 30 years [12].

To close the debate on multicore, Baek and colleagues argue that, after a series of technical innovations, the existing semiconductor technology is now at its technical limit [13]. They further argue that especially the processing speed of the CPU is now even with the physical limit which is the speed of electrons [13]. Thus, in order to overcome this situation, a new paradigm is needed. Throughout the previous sections, we have tried to explore some paradigms for models of parallel computation. The objective was to show that there have been quite a considerable number of models developed, but the result has not been entirely satisfactory. In the next subsection, we look at some other attempts carried out for the same purpose.

### 3.5. Other Parallel computational Models

Apart from the approaches discussed previously, there exist some other models in parallel computing that we can mention in this research. As with the attempt of the last PRAM variant discussed earlier in this text, the DRAM, created to alleviate the concept of global shared memory, another paradigm in this area was already triggered where memory modules are linked to clustered processors [63]. An illustration of such models is the Distributed Memory Model (DMM) [63]. In this model, private memory modules are associated with processors in a bounded degree network [46]. In this model, a time step is used to coordinate the computation, as well as communication between neighboring processors.

Another parallel computational model that was built around the concept of memory modules is the Postal Model [14]. The Postal Model operates in the same way as an ordinary mail system works. In order to exchange information or transfer data to a non-local memory, a clustered processor posts the message into the network and the latter takes care of the rest [46]. Moreover, the Postal Model is one that puts a critical emphasis on parallel communication. As its name suggests, the model deals more with how communication occurs within the architecture and no further description of the computation mechanism is provided. The Postal model provides a unique and specific mechanism that caters for communication latency. Its variant called the Atomic Model for Message Passing was developed to include a bit of computation, even though the implementation of the latter carries with it a specific cost. This cost is related to the complexity of its configuration and implementation [45].

Early models in the context of serial computing and sequential hierarchical memory include, for example, the Hierarchical Memory Model (HMM) and Block Transfer (BT) [7] models. In these models, particularly the first model, access to memory occurs through memory locations that are contained on each level of memory in the model. The second model, however, is a more sophisticated and more improved version of the first one that makes it possible to move data in larger chunks or blocks [46].

While the output or results of such models in serial computing are significant, as they allow the movement of larger amounts of data across the intertwined communication links, they could not be used as appropriate or reference design tools.

Hence, the parallelised versions of these two models were developed to try to see the possibility of successfully modelling data access in parallel architecture. These models include the parallel HMM (PHH) and parallel BT (P-BT) . The refinements in the parallelised version of these models include the replication of serial model execution times and the connection of processors and memory via a network that will serve as a common infrastructure to allow the concurrent and parallel movement of data. Though these models provide significant performance results such as the ability to exploit both parallel and block transfer of data in communication channels in relation with data movement, they still use sequential or hierarchical execution by refining the existing sequential models. Hence, this approach becomes costly in terms of memory size and access for each configured memory level [46].

## 4. NEED FOR A NEW MODEL OF PARALLEL COMPUTATION

The need for this new model dates from a number of decades and is heightened by many factors. One aspect is the greater demand for performance and the greater diversity among current machines. An additional reason is the need to allow the translation of algorithm steps into a sequence of instructions that are machine-independent [48] and [46]. The model will guide the implementation of an appropriate programming model. Specifically, a successful computational model would allow the operation of a programming model that provides a collection of rules determining the significance of programming tasks expressed through a programming language as the model instantiation [46].

Maggs and his colleagues emphasised that the primary focus in the process of programming languages development has been for many years based on providing functions as well as constructs whose objective is to translate algorithmic intentions [46]. However, the development of any programming language also relies on the translation of functions and constructs of high-level language into executable instructions that are dependent on executing machines. This translation is hence facilitated by the use of an effective and appropriate model of computation on which the machine architectural design is based. Therefore, in parallel computing there is an increasing need and demand for high performance not only in computing machines but in language development as well. The development of programming languages as well as many other computer applications in parallel computing requires an appropriate parallel architectural model.

On the other hand, performing very high-computational problems requires an architecture that is based on and, very importantly, designed according to the HPC (High-Performance Computing) framework [12]. Such architectures consist fundamentally of clustered parallel processors, thousands of CPUs [58]. High Performance Computing creates new demands on software applications with respect to performance, scalability and portability. The increased complexity of parallel machine architectures, on one hand, and different parallel programming paradigms on the other hand, is alarming as to the need for a new and more reliable computational architecture that would successfully bridge this gap.

Additionally, a new model will help provide a bridge between hardware and software to assist application developers in designing and writing parallel applications that efficiently exploit parallelism [58]. However, many existing architectures have not been able to meet this requirement and produce optimal and permanent solutions. Also, the von Neumann machine gives poor results in a multiprocessor configuration because it fails to provide efficient synchroniation support at a low level [37].

Furthermore, the Berkeley group believes that the new model would be one that embodies standard parallel performance characteristics including simplicity, naturalness and, more importantly, efficiency [12]. In their analysis of parallel computing research, Asanovic and Colleagues (Berkeley group) support that parallel computing models should be more human-centric than the conventional focus on hardware or applications if they are to be deemed successful. However, current parallel architectures strongly rely on the random access machine approach. These models fail to maintain consistency throughout the development and implementation of computing architectural designs, given the complexities in their structure as well as their performance constraints [12].

The conducted analysis over these models would suggest that, taken in isolation and individually, none of these models seems to incorporate all the necessary characteristics for a robust, efficient and acceptable parallel computational model that could be used as an indicative design tool. However, when considered as a group, the ultimate characteristics of a unique and appropriate model seem to be found on the different isolated models. Throughout this paper, it has emerged that most of these models attempt to model few characteristics or requirements of machine design individually. These requirements include, for example, communication overload, latency, parallel execution coordination and synchronisation, communication and bandwidth, as well as performance [46]. Therefore, the new model is one that will combine all these factors and handle parallelism efficiently in a cost-effective fashion while enabling high performance computation.

## 5. DISCUSSION AND CONCLUSION

In this paper, we have attempted to summarise and review related work in terms of parallel processing. From exploring the importance of parallel processing and modelling in computing, the paper also gave an overview of some efforts undertaken in order to develop an appropriate model of parallel computation. Some existing approaches including the PRAM model, Dataflow model, BSP model, the multicore paradigm as well as other models were briefly discussed to give an idea of existing work in the field and the need for a new model. This paper has also tried to demonstrate that it is critical and of great significance that a new model of parallel computation model be considered to meet the demand for high performance computing.

Lastly, we tried to demonstrate in this paper that many research studies indicate that parallel models such as PRAM that is based on the von Neumann model still need more improvement because of a number of factors ([46]; [36]). These factors include weaknesses in specifying computation as a sequence since the semantics of memory is too complex and this results in many errors. A memory location has many meanings. Another factor is that there is little control over the sequence of instructions because of issues such as a jump to a random address (Mueller, 2009). The last factor is a lack of, or little control over, the access and modification of memory addresses. This may result in exceeding array bounds ([46]; [36]). All these factors are reasons that motivate the need for a new model of parallel computation. This research study has been triggered by this need to consider another candidate for this purpose. This candidate and new model is the AriDeM architectural model.

A detailed description of the model can be found in [51]. In [1] and [2], initial evaluation and performance results of the model are presented. The main objective of this research was to evaluate AriDeM as a new model of parallel computation. Before starting the actual evaluation, we attempted to give an overview of the model. AriDeM is a model of parallel computation that is based on natural computation rules. It processes elements as opposed to instructions being executed in the von Neumann model. One of the main features that we discussed was the ability of processors to process elements independently of others. This increases parallel performance as it eliminates communication overhead. All of this needed to be evaluated to help determine whether AriDeM is a suitable candidate for the needed new model of parallel computation.

To perform this evaluation, one would look at the performance of AriDeM in the light of how the existing approaches perform. Matrix multiplication was chosen as the case study to be used for this evaluation. The evaluation process occurred twofold: theoretical analysis and empirical study. Moreover, during the evaluation, two main metrics were used: the execution time to help depict the scalability for both models and the number of instructions executed compared to the number of elements processed. This helped show which architecture was likely to reduce problems affecting performance such as communication overhead etc. The number of executed instructions and processed elements would also give an indication of an approximate number of memory accesses, which in return constituted an additional metric used in this research.

In theoretical analysis, the core objective of the theoretical evaluation was to demonstrate that the element model performs as well as the von Neumann model on a single processor and that in parallel it actually gives insight into producing better results than the instruction model can. Hence, to prove this theory, we looked at the high-level program in assembly language for both models. From the evaluation, it appeared that, in terms of verifying whether executing an instruction is equivalent to processing an element, the results obtained from the execution of the code established this assumption. The evaluation also showed that, from the performance in terms of the number of instructions executed for the von Neumann model and the number of processed elements for the element model, the latter is more effective. The analysis gave an indication that fewer elements were processed to accomplish the same tasks in the element model, as compared to the number of instructions executed to perform the same task for the instruction model.

With regard to the parallel versions of both programs, the compiled results of these two programs showed how the models compare for the parallel versions. The results showed that the computation of large matrices multiplication requires a relatively greater amount of time, as it is of *O ($n^3$)* order of complexity. Hence, for the execution of this algorithm in parallel, the parallel algorithm required *$2*n^3$* arithmetic operations. Hence, the optimal parallel time on p processors will be *$2*n^3/p$* time steps. Looking at the results of the parallel versions of the program for both models, a significant feature of the element model was noticed: whilst a complete rewrite of the code for the von Neumann model was required to parallelise the program, the simulation of the AriDeM architectural did not, on the other hand, impose this burden. This is important, since it shows that the model handles parallelism at both the hardware and software level.

The empirical study provided proof of the results estimated from the theoretical evaluation. We looked at empirical evidence to help determine the scalability for both models. This evaluation also helped get an indication of how AriDeM would perform, since we have been using its simulator. We created two separate programs in order to run a matrix multiplication program for both models. The experiment proved that a program like matrix multiplication can be written in the paradigm of the proposed model that can run on different configurations without any change and running it on *2, 4, 8* and *16* processors gives the same results.

Despite the algorithm complexities, memory latency and other factors affecting the overall performance of the architectures, the results were enough to support the first findings and observations. The experiment gave an indication of how scalable these two models are. The instruction model could not show an effective synchronisation of data processing as the results do not indicate scalability easily. Looking at the different graphs in this paper, we can depict some discrepancies in terms of execution time and increasing size of matrix. We notice that as the size of matrix increases, the model does not strongly express that it is scalable. On the contrary, when the matrices *80 by 80* and *100 by 100* and even *60 by 60* are run on *8* and *16* processors, we notice that the time actually goes up with 16 processors instead of decreasing. The same happens slightly for AriDeM but only with the *60 by 60* matrix. Otherwise, AriDeM distinguished itself as being just a bit more scalable than von Neumann in this case.

In an attempt to get some comparison with a conventional architecture, we compared the number of instructions that need to be performed on the instruction architecture with the number of elements that needed to be processed on this architecture. For the same matrix size, in order to do the same thing, AriDeM processes fewer elements compared to the number of instructions that are executed by the instruction model. As the graphs indicated, the number of elements processed in AriDeM is almost three times fewer than the number of executed instructions with von Neumann. These results also helped understand other aspects of the model compared to other approaches.

The comparisons with other approaches show that the proposed model is significantly different from models currently in use. The differences include the form of expressing programs, the way programs are evaluated, and the stricter form of referential transparency etc. This experiment has proved that having such a different model of computation is of value in its own right. It gives a different perspective on how computation can be done.

The nest step will socus on conducting more experiements with AriDeM with different case studies in order to give more evidence that would enable one to draw more conclusive and mature remarks although at this stage the prospects show positive results.

**BIBLIOGRAPHY OF AUTHOR**

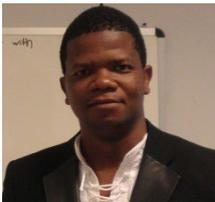

Patrick Mukala is currently a PhD Candidate in Computer Science at the University of Pisa in Italy. This work was done as part of his Masters of Technology in Software Development at the Tshwane University of Technology.